# Enhancement and Inhibition of Transmission from metal gratings: Engineering the Spectral Response

D. de Ceglia[1], M. A. Vincenti[1], M. Scalora[2], N. Akozbek[2], M. J. Bloemer[2]

[1]*AEgis Technologies Inc., 401 Jan Davis Dr, Huntsville – AL, 35806, USA*
[2]*Charles M. Bowden Research Center, AMSRD-AMR-WS-ST, Redstone Arsenal, Huntsville – AL, 35898, USA*

**ABSTRACT**

We present a systematic analysis of the optical properties of slit arrays in metal films. An exhaustive investigation of geometrical and dispersive properties reveals the resonance features of these structures, including the role of surface waves and their relationship with features in the transmission spectrum. Although enhanced transmission windows are significantly dominated by the longitudinal resonances localized inside the slits, the periodicity introduces transverse resonances that can either enhance or inhibit light transmission. We thus illustrate the intriguing interaction regime between longitudinal and transverse resonances, where the two modes hybridize leading to the formation of a photonic band gap spectrum.

**INTRODUCTION**

Since the first observation of unexpectedly narrow bright and dark bands in the reflection spectrum of optical gratings in the early 1900s [1, 2], many efforts have been devoted to the explanation of these anomalous phenomena. The first theoretical interpretation of these effects, which could not be explained by means of ordinary diffraction-grating theory, was provided by Lord Rayleigh [3], and subsequently refined by Fano [4]. In particular, Lord Rayleigh's theory predicted the spectral positions of the anomalies resulting from the appearance of new spectral orders. A complete and fascinating interpretation of Wood's anomalies was reported many years later by Hessel and Oliden [5]. They elucidated two different anomalous manifestations of the grating: (i) the abrupt intensity modulation of diffraction orders at the appearance or fading of new spectral orders, and (ii) resonant-like anomalies. These two effects can occur either separately or simultaneously. In particular, the resonant effect corresponds to the excitation of leaky waves supported by the grating for specific spectral orders. The possible role of surface plasmons in the formation of these anomalies was discussed for the first time only after 1968, when Otto [6], Kretschmann and Raether [7] reported the excitation of surface plasmons on continuous metal films.

In 1998 Ebbessen et al. [8] demonstrated the extraordinary optical transmission of light through metal layers perforated with cylindrical, sub-wavelength holes. In explaining the effect, the role of surface plasmon polaritons was emphasized. The spectral proximity of Wood-Rayleigh transmission minima and surface plasmon modes has made the physical interpretation rather difficult, and has given way to contrasting views. Moreover, several authors have reported extraordinary transmission from one-dimensional (1-D) sub-wavelength slits instead of holes, and in different regimes, from the microwave to the UV range [9-14]. However, what goes on inside a narrow slit or slits is quite different from occurs inside cylindrical apertures. The former is strongly dominated by the transverse electric magnetic (TEM) waveguide mode propagating in the longitudinal (or horizontal) direction. In contrast, the latter does not support TEM modes.

Hereafter, we use the terms horizontal or longitudinal to indicate the direction perpendicular to the free-space/grating interfaces, while the terms vertical or transverse designate the direction parallel to the interfaces. A 1-D metal grating composed of rectangular slits supports wide-band Fabry-Perot-like resonances whose material and geometric dispersion can be significantly altered by the presence of the grating. The main effect of the periodicity is, indeed, the activation of guided modes along the vertical direction, whose excitation would not be possible without evanescent coupling mechanisms (i.e. a typical prism coupling). The additional wave-vector introduced by the grating along the parallel direction, $k_G = \frac{2\pi}{P}$, where $P$ is the pitch size, induces guided modes on the grating located above the light line, and eventually opens transmission resonances in the normal direction.

During the past decade two contrasting characters have been attributed to surface plasmons in metal gratings. The first view links transmission maxima to the excitation of surface modes favored by the presence of the holes/slits [15-17]. The second view, relying also on experimental evidence, imputes a mere negative role to surface plasmons [18-20]. Authors supporting the positive influence of surface plasmons on the extraordinary transmission process usually attribute the transmission minimum to the Wood's anomaly [15] and explain the discrepancies between the predicted and calculated (or measured) maxima as the effect of the perturbation induced by the modulated surface [21] on the surface plasmon mode. The dispersion relation is commonly defined at an interface between a dielectric material and a smooth metal surface. Under this point of view, corrugations, grooves, holes and slits just barely shift the dispersion relation of the surface mode as defined in [22]. On the other hand, opponents of this school ascribe to the combination of diffraction and interference of modes the origin of extraordinary transmission [18], while surface plasmons are confined to have only a negative role in the transmission process [18, 20, 23].

These opposing views can lead one to ask the following simple question: what is the actual role that surface plasmon polaritons play in the enhanced transmission process? In order to answer this question in a satisfactory way one should start at the beginning with the definition of what is surface plasmon polariton. Most authors would agree that a surface plasmon is a collective, resonant oscillation of charges induced at the interface between a positive and a negative permittivity material [22, 24]. Using this rather unambiguous definition, the dispersion of this surface mode is well-known and given by $k(\omega) = \frac{\omega}{c}\sqrt{\frac{\varepsilon_1 \varepsilon_2}{\varepsilon_1 + \varepsilon_2}}$, where $\varepsilon_1$ and $\varepsilon_2$ are the permittivities of the adjacent materials. In the terahertz and microwave regimes this dispersion curve nearly overlaps the light line ($k = \omega/c$). At infrared wavelengths and beyond the curve is well detached from the light line as it bends toward the surface plasmon resonance $\omega_{sp} = \sqrt{\frac{\omega_p}{1+\varepsilon_2}}$, where $\omega_p$ is the plasma frequency of the metal that supports the surface plasmon propagation.

The excitation of this particular mode plays an important role in the analysis of transmission and reflection spectra of periodic arrangements of sub-wavelength slits both in free-standing or supported metal layers. On the basis of recent theoretical and experimental work [18-20], confirmed by the results presented here, the grating becomes a virtually perfect reflector whenever this mode is excited. In fact, as first observed by Wood in 1902 and clearly explained by Lord Rayleigh in 1937, a strong modulation of the transmission efficiency occurs at the onset, or disappearance, of new diffraction orders, whose wave-vector must be matched to a grazing mode. The surface plasmon polariton is itself a resonant part of the diffracted energy [25], and as such it may also be interpreted as a particular case of this mode, and should be studied in relation to all the phenomena involved in the diffraction process. As a matter of fact, the characteristic wavelength of this mode, as well as its effective refractive index dispersion, may be altered by the geometric parameters of the grating, such as aperture size, thickness of the metal film, and grating pitch. As demonstrated by Pendry et al [26], the dispersion of surface waves may be engineered by properly structuring the surface, even in the presence of perfect metallic screens.

In this paper we report a comprehensive analysis of the optical response of metal gratings focusing on symmetric, rectangular slits carved on silver layers without loss of generality. The role of each of the geometric parameters involved in the process is studied and explained. In particular, we show that the inhibition of transmission occurs when the impinging parallel wave-vector, added to the grating lattice wave-vector, matches the surface plasmon wave-vector of the unperturbed air-metal interface. We are mostly concerned with incident TM-polarized fields, so that the sub-wavelength sized geometries do not undergo any cutoff effect, at least for the TEM mode. The case of TE polarized light (having the electric field only tangential to the interfaces) shining onto an array of sub-wavelength slits sandwiched by two dielectric layers is also presented, demonstrating how enhanced and inhibited transmission can be mediated by grating resonances, even though surface plasmon modes may be absent. Furthermore, we explain the difference between surface plasmons and Wood's anomaly by studying the transmission response of gratings with variable aperture sizes. The tools for engineering the spectral response of such structures are provided, showing the dynamical interaction of longitudinal and transverse resonances. In fact, the interactions between longitudinal modes confined inside the slits and transverse modes propagating along the grating allow the formation of plasmonic [27, 28] or photonic band gaps. A generalization of the Rayleigh minima wavelengths and the Wood's anomaly will be proposed and discussed, that can be straightforwardly applied to both transverse electric (TE) and transverse magnetic (TM) polarizations.

**FROM A SINGLE SLIT TO THE INFINITE ARRAYS: INFLUENCE OF THE GEOMETRIC PARAMETERS ON THE TRANSMISSION RESPONSE**

In order to capture all the aspects of the phenomena that take place in sub-wavelength sized metal gratings, we begin by analyzing the problem of a single rectangular slit carved on a silver substrate. All calculations and results reported in this paper were obtained using at least three different computational methods that yield nearly identical results: a commercial code based on the Finite Element Method (Comsol Multiphysics); a 2D-FDTD and a time-domain FFT-BPM whose details have been discussed elsewhere [29, 30]. Some slight differences are only observable in proximity of the Wood's anomaly, where the use of very long pulses is required to capture the extremely narrow spectral features around these regions. The dispersion profile of silver is the one found in Palik's handbook [31]. The influence of the geometric parameters, i.e. slit size and film thickness, as well as their respective dependence to the impinging wavelength, have been considered in attempt to extract as much information as possible about the nature of the interaction between the geometrical features and the transmission response. As shown in Figs. 1 (a), (b) and (c), a single slit system exhibits pronounced resonant behavior whose nature is very similar to that of Fabry-Perot resonances. In this relatively simple system, the guided modes (mainly the TEM mode, along with the plasmonic modes of the metal-air-metal waveguide) are poorly coupled to free-space, except at the resonant wavelengths $\frac{m\lambda_0}{2n_{eff}}$.

The transmission peaks of these resonances are governed by both geometrical parameters, which are thickness of the metal layer $w$ and slit size $a$: Fig.1 (a) is obtained by considering an incident plane wave tuned at $\lambda = 600$nm impinging at normal incidence onto a grating with variable slit size and film thickness. The values of *a* and *w* are tuned in the ranges 20nm÷300nm and 50nm÷600nm, respectively, and the transmittance is normalized to the portion of energy illuminating the slit.

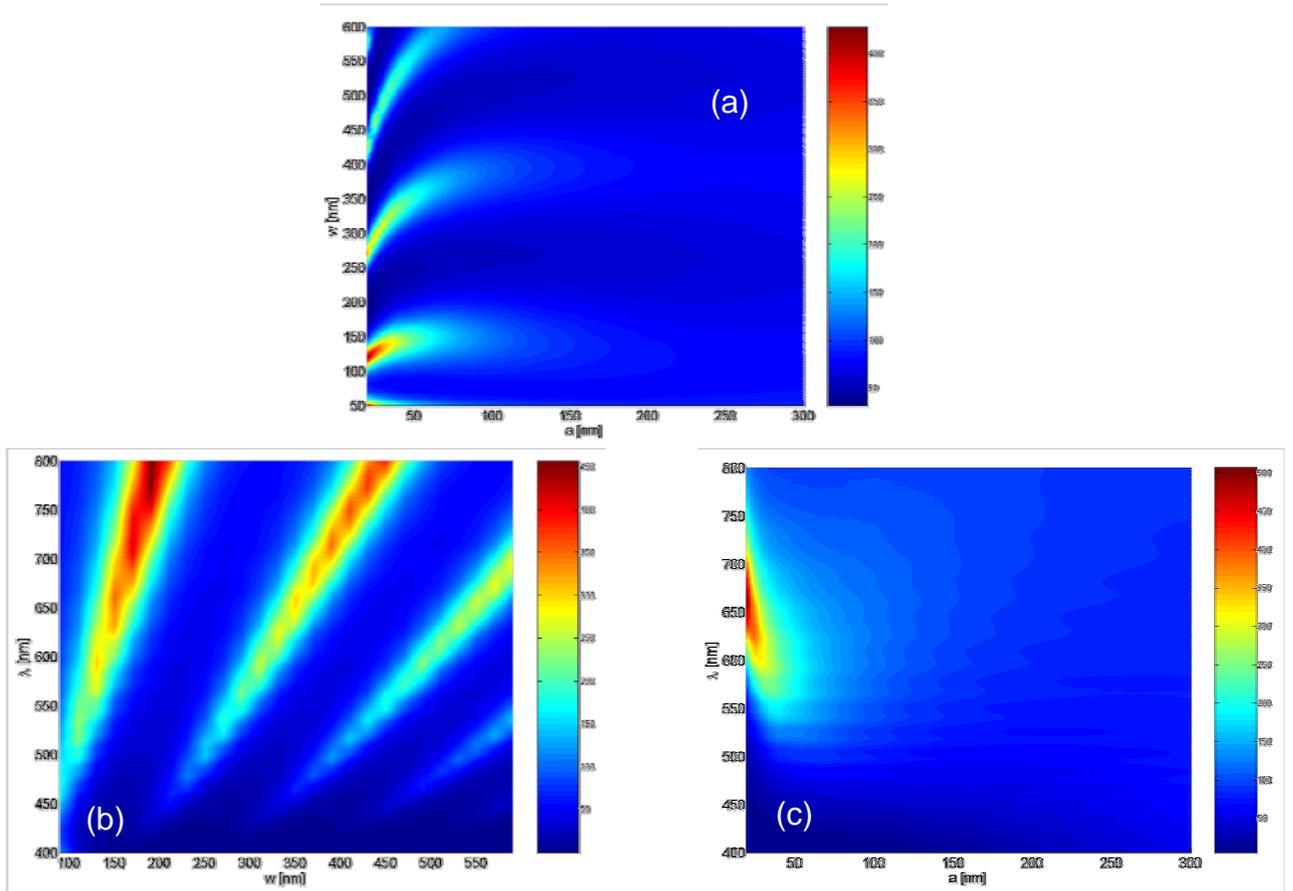

**Fig.1:** Two dimensional transmission maps for a single slit carved in a silver film. Three free parameters have been varied in turn; (a) transmission coefficient for an impinging wavelength λ = 600nm and variable *a* and *w*. (b) transmission coefficient for a variable incident wavelength and varying *w*; *a* = 32nm; (c) transmission coefficient for a variable impinging wavelength and varying *a*; *w* = 135nm.

As already pointed out elsewhere [14, 16, 32], these kinds of resonances are characterized by significant field localization inside the isolated slit and they induce enhanced transmission without any contributions from surface corrugations or any grating effects. The extensive spectral modulations reported in Figs.1 result from scanning several horizontal resonances and give a clear idea of how high the transmission from a single sub-wavelength slit can be. In a single slit, in fact, the transmission process is mostly driven by the TM fundamental mode, whose dispersion is strongly influenced by metal conductivity and slit size. If one considers a waveguide made by two parallel plates separated by a free-space gap, the effective index of the fundamental TM mode is exactly equal to one only in the case of a perfect electric conductor (PEC): in that case the field profile inside the gap region is constant, and zero in the metal.

The introduction of a finite conductivity allows the mode to penetrate inside the metal and hence cause an increase of the effective index. This means that the dispersion of this modified mode follows the dispersion of the metal. Moreover, even at microwave wavelengths, where the conductivity is very high, if the sub-wavelength gap is small enough it will show an effective index slightly higher than unity [33]. To understand this phenomenon it is enough to keep in mind that the effective index of a mode propagating in a waveguide is a weighted average of the permittivities that constitute the transverse section of the waveguide. The weights are proportional to the mode overlap in each region. Hence, if the gap is much smaller than the electromagnetic wavelength, the small portion of modal energy propagated by the metal cannot be neglected with respect to the fraction of the mode travelling inside the gap.

Another effect that should be considered when the waveguide is abruptly truncated is the strength of the coupling between the guided mode and free space. Even when the surrounding metal is assumed to be a PEC, so as to have an effective modal index equal to one, at the exit of the waveguide the mode undergoes a reflection coefficient quite far from the π-phase shift ordinarily introduced by a perfect mirror, as in an ideal Fabry-Perot interference system. In fact, coupling to free space could be improved using any of a number of strategies, for example, by tapering or rounding the entrance and exit apertures of the slit in order to improve the diffraction efficiency and reach a better matching between the free-space mode and the waveguide mode. As a result, the effect of truncation is a spectral shift of the horizontal resonances, which should be added to the above mentioned shift induced by the finite conductivity of metal.

The existence of the main propagating mode supported by the parallel-plates waveguide, which is indeed available without cutoff only for TM polarization, leads to a significant enhancement in the transmission that approaches values as high as 500% with respect to the energy that impinges on the geometrical area of the isolated slit. Strongly resonant behavior is evident in Fig.1 (b) and (c) as well. The figures were obtained by varying only one of the geometrical parameters mentioned above ($w$ in Fig.1 (b) and $a$ in Fig.1 (c)) and the impinging wavelength, that varies in the visible range. The other geometrical parameter has been arbitrarily fixed to $a = 32$ nm (Fig.1 (b)) and $w = 135$ nm (Fig.1 (c)).

A multiple slit geometry structured in a one dimensional periodic array introduces the pitch of the resulting grating as an additional degree of freedom. It is well known that adding more slits does not enhance significantly the transmission value itself, which indeed tends to saturate for more than 6 slits [20, 32]. This argument holds only when the pitch is smaller than the resonance wavelength. If pitch size is close to the resonance wavelength, or inside the high-transmission band, the spectral response can be altered significantly. As an example, in Fig.2 we show the transmission spectrum at normal incidence through a single slit (blue line - circle markers) compared to the transmission spectra of multiple-slit arrays having increasing number of slits, and to an infinitely long array (dark blue line - cross markers). Aperture size is fixed at $a = 32$nm, the silver thickness is chosen $w = 200$nm, and the pitch size is $p =$ 820nm. In order to understand the effect of multiple slits our starting point is with a free-standing grating surrounded by air. In this case at normal incidence Rayleigh minima occur around the wavelengths $\lambda = p/n$, where n is an integer. As a consequence, by choosing a pitch that falls inside the Fabry-Perot-like resonance we are introducing interference between the Wood's anomaly and the horizontal resonance of the system. In going from one to two slits the resonance narrows and the maximum transmission increases. With just a few more slits the Wood's anomaly effect becomes evident, so that the Fabry-Perot resonance undergoes strong reshaping driven by the periodicity of the array.

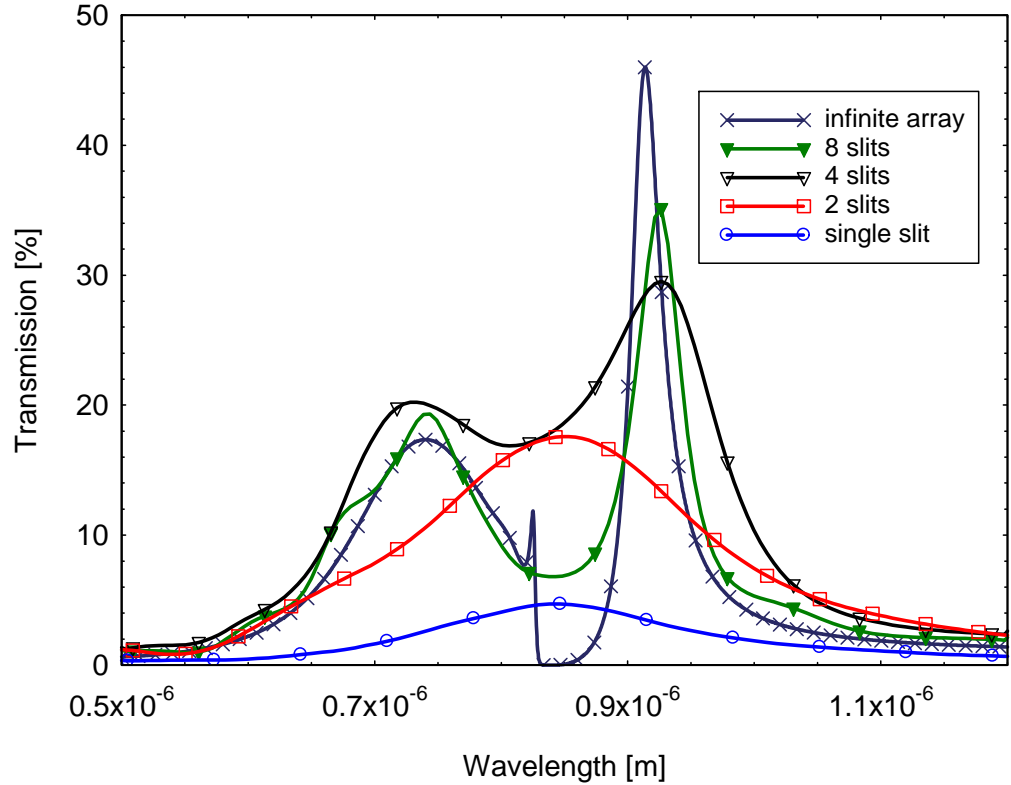

**Fig.2:** Comparison of transmission spectra at normal incidence showing the formation of a band gap related to the array periodicity: a single slit (blue line, circle markers) shows the same Fabry-Perot feature of the array, being the resonance related only to the slit size and film thickness. Increasing the number of slits to 2 the spectrum shape does not change and the transmittance increases (red line, square markers). With 4 (black line, empty triangle markers) and 8 slits (green line, full triangle markers) a dip appear at the Rayleigh minimum position, which is more pronounced for increasing number of slits. In the limit of an infinite array the dip turns into a band gap (dark blue line, cross markers).

The interference between propagating and evanescent orders diffracted by the grating induces a split of the original resonance with the introduction of a minimum at the Wood's anomaly position. The effect becomes even more pronounced in the limit of an infinite number of slits (dark blue line, cross markers), for which the dip becomes a wide gap and a sharp transmission edge appears at a slightly blue-shifted wavelength. As also happens with structures that are periodic on the scale of the wavelength of the impinging light, the addition of more periods in the structure favors the opening and widening of a band gap, whose center and width is mainly controlled by the periodicity of the structure.

The similarity of a surface plasmon travelling on a corrugated surface and a photonic band gap structures, as well as the relationship between the band gap and the periodicity, have been already demonstrated for shallow and deep grooves on metal films [27, 28]. Two well-defined states are separated by an energy gap, i.e. a plasmonic band gap, in which wave propagation is forbidden along the surface. Even in the diffraction grating that we are considering the formation of these gaps in the spectrum is mainly controlled by the pitch size. In an infinite array of slits, there are three additional mechanisms that intervene that are absent in the single slit case: (1) the interaction between diffracted waves; (2) the introduction of surface waves triggered by the reciprocal lattice vector of the grating; (3) the perturbation induced by the grating itself on surface waves.

The simultaneous occurrence of these effects and the proximity of the Wood's anomaly to the conditions for the excitation of an unperturbed surface plasmon on the metal surface make it especially difficult to fully discern the complex physics of this apparently simple system. In the past, some authors have claimed that the primary role of surface plasmons is to mediate the enhanced transmission process. Others have argued and demonstrated instead that the surface plasmon excitation corresponds to the formation of transmission minima. In our view, these contradictions probably derive from early attempts to simplify the first observation of enhanced transmission in nano-hole arrays with the study of periodic arrangements of slits under TM polarization. In the former, vertical resonances (surface waves, leaky waves, plasmon polariton modes) are the only possible vehicles able to catalyze evanescent waves to channel part of their energy through the holes, even under cutoff. In slit arrays, the horizontal resonances are the dominating transmission mechanism, and surface waves can only interact with Fabry-Perot-like modes, forcing at times severe reshaping of the transmission spectrum near the Wood's anomalies. We will discuss also the complexity of this interaction, which leads to the formation of the gap/edge structure reported in Fig 2.

For the sake of simplicity from now on we will focus only on infinite arrays. These structures fully preserve the Fabry-Perot resonance positions, the characteristics of their single-slit's counterpart, and remain dominant features. However, by varying the pitch size $p$ one can either isolate these Fabry-Perot modes, or cause them to interact with the diffraction process taking place on the grating. In Fig.3 (a) we plot the transmission of a single, rectangular slit with an aperture size $a = 32nm$ and a length $w = 300nm$ that pierces the silver layer from side to side. In this spectrum two resonances can be easily recognized around 550nm and 1100nm. If one considers an array of these slits with a pitch size $p = 280$ nm, the resulting structure is in the zero-order grating condition

$$\frac{\Lambda}{\lambda} = \frac{1}{n_{in}\sin\theta + \max(n_{in}, n_{out})}, \qquad (1)$$

where θ is the incident angle and $n_{in}$ and $n_{out}$ are the input and output refractive indexes, respectively. As a result, the grating is sub-wavelength and the only forward-propagating diffracted wave is the zeroth order. All higher orders are evanescent. As displayed in Fig.3 (b), there are no significant modifications in the shape of the spectrum, which matches quite well the resonances of an isolated cavity of the same size.

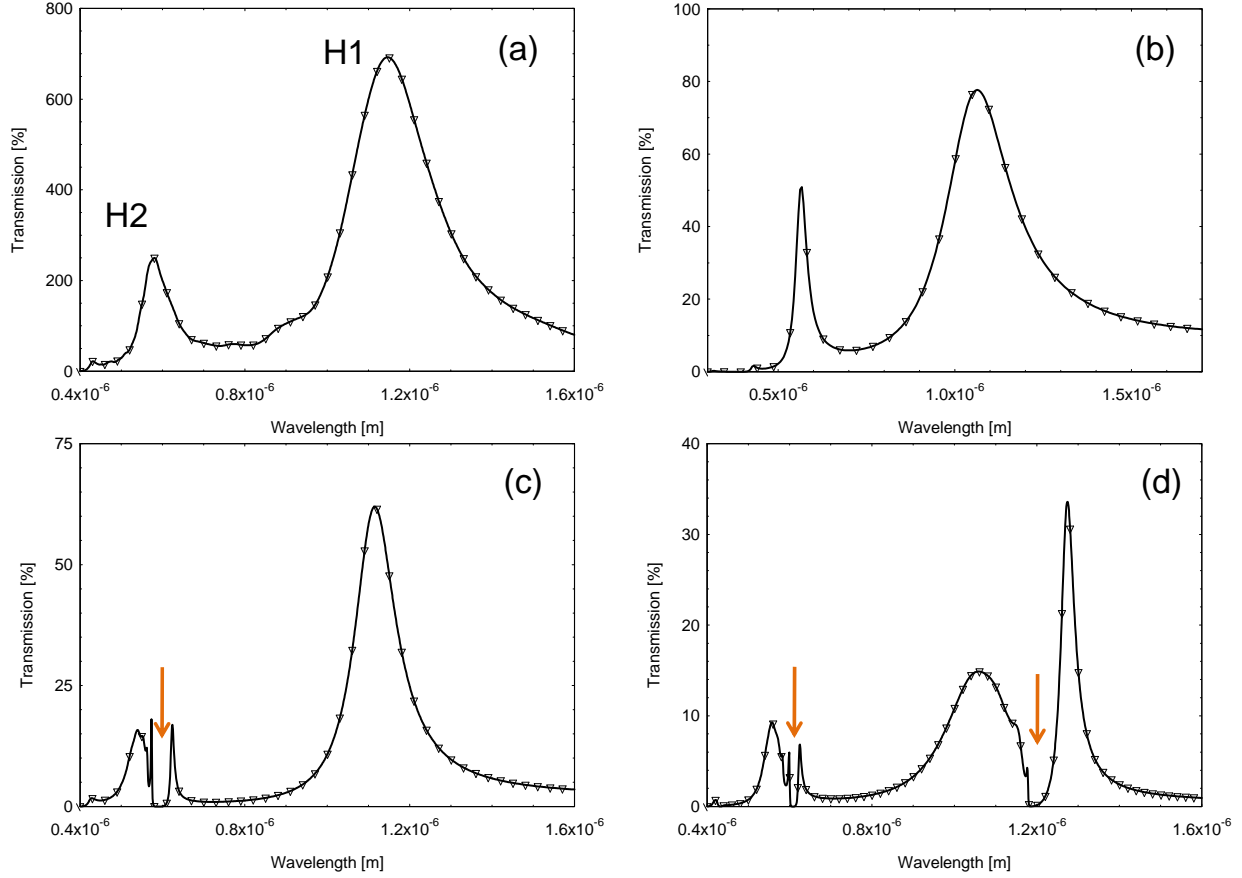

**Fig. 3:** (a) Transmission spectrum of a single 32nm slit in a 300nm silver film, normalized with respect to the energy incident on the geometrical area of the aperture. The transmission spectra at normal incidence for silver gratings with different pitch sizes are reported in (b), (c) and (d), normalized with respect to total incident energy. The geometrical parameters for the gratings are *a* = 32nm, *w* = 300 nm and (b) *p* = 280nm, (c) *p* = 566nm, (d) *p* = 1174nm. Figures (b), (c) and (d) show the formation of (b) 0, (c) 1 and (d) 2 band gaps in the spectra, caused by the periodicity of the array.

The reason for this derives from the simple observation of the position of the Wood's anomaly wavelengths, $\lambda_W = \left|\frac{\sin\theta}{\lambda_{0,inc}} + \frac{m}{p}\right|^{-1}$, where $m = \pm 1, \pm 2, ...$, θ is the angle of incidence and $\lambda_{0,inc}$ is the incident wavelength. We emphasize that the horizontal cavity modes of the single slit, named H1 and H2 in Fig. 3(a), guarantee a total transmission of 80% and 50% (Fig.3(b)), respectively, even with a metal film 300nm-thick, which in this operating regime is virtually opaque without apertures.

On the other hand, by tuning the period of the grating inside the resonance bandwidth one is able to drastically alter the transmittance function, and to reshape one or more Fabry-Perot resonances simultaneously. As Figs. 3 (c) and (d) show, a periodicity $p = 580$ nm or $p = 1174$ nm splits the horizontal resonances and creates one band gap (Fig. 3(c)) and two band gaps (Fig. 3(d)) respectively. The effect of this interaction manifests itself with the near-complete inhibition of transmission when the incident wavelength excites the surface plasmon of the unperturbed air-metal interface. This condition takes place for those surface plasmon wavelengths that match *exactly* the grating pitch, so that at normal incidence $\lambda_{spp} = p/m$. A careful look at the transmission spectrum is in fact enough to notice that for very small apertures the excitation of the surface plasmon makes the grating a very efficient reflector, as the apertures are located at positions that describe minima of the electromagnetic field. In this situation, the impinging light exploits the sub-wavelength features of the grating to find the missing momentum that allows the propagation of a surface plasmon without resorting to any other evanescent coupling mechanisms. On the other hand, since the aperture is significantly smaller than the incident wavelength, the surface wave propagating on the air/metal interface has a dispersion relation that is not significantly modified with respect to the surface plasmon of an unperturbed metal layer.

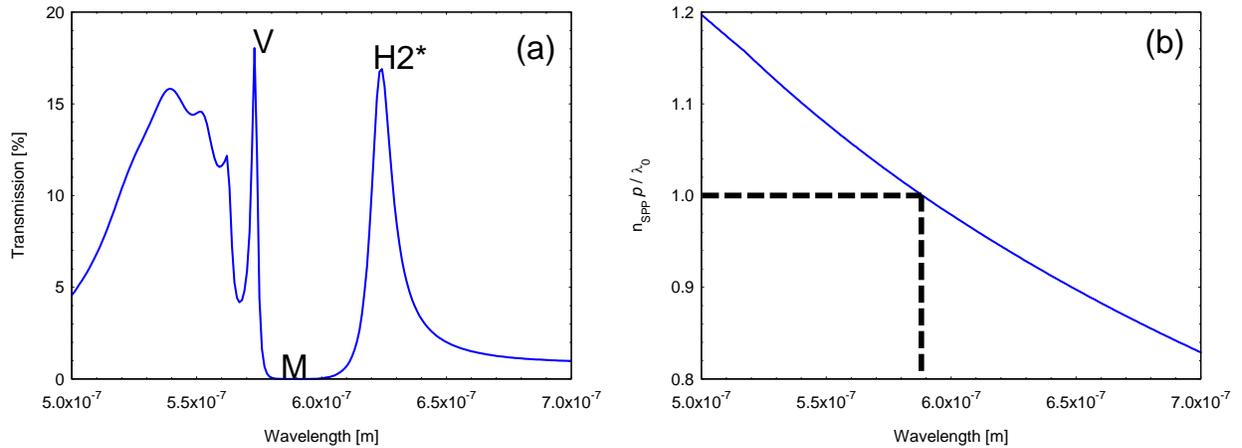

**Fig. 4:** (a) Transmission spectrum at normal incidence of an array of sub-wavelength slits with the following geometrical parameters: $a = 32$ nm, $w = 500$ nm, $p = 566$ nm; (b) Surface plasmon polariton wavelength in normalized units plotted as a function of the incident wavelength. The highlighted point represents the condition for the excitation of the unperturbed surface plasmon polariton, which leads to a formation of the transmission minimum M reported in Fig. 4 (a).

To provide further evidence that the minimum is produced by the excitation of a surface wave along the air/metal interface it is important to investigate the field localization at the transmission minima across the spectrum. For this purpose, let us consider the spectrum illustrated in Fig.3 (c), which refers to a grating pitch $w = 300$ nm. The horizontal resonance (H2) around 550nm (Fig. 3(a)) interferes with the surface wave excited in the vertical direction, while the resonance around 1100nm (H1) is only slightly perturbed by the presence of the grating, in that its bandwidth is slightly narrowed with respect to the single-slit spectrum reported in Fig. 3(a). A magnification of the spectrum around the H2 wavelength is reported in Fig. 4(a). One may recognize a sharp resonance at $\lambda_v$= 573 nm (denoted as the V resonance in the figure), a gap located in the neighborhood of the unperturbed plasmonic wavelength $\lambda_0 = n_{spp}p$ , and two wider maxima on each side of these anomalous spectral features (the V maximum and the gap introduced by the periodicity), which are reminiscent of the horizontal resonance. We name H2* the hybridization of the horizontal resonance H2 (excited by the single–slit) with the vertical resonant phenomenon triggered by the diffraction grating. The last point marked in Fig. 4(a) is the minimum M, whose nature, as discussed above, is strictly related to the excitation of the unperturbed surface plasmon polariton. The location of M ($\lambda_M$ = 588nm) is explained in Fig. 4(b), where we provide a graphical solution of the equation $\lambda_0 = n_{spp}p$ for the grating under consideration.

The field distribution depicted in Fig. 5(a) refers to the point H1* of Fig. 3(c), where the only diffracted wave is the zero-order, and the grating is far from the conditions necessary to excite vertical resonances. This lack of vertical resonance is confirmed by the frustration of the magnetic field at the input and output interfaces and the absence of any significant perturbation of the Poynting vector, even in the near-field of the aperture. In this case one may conclude that, from a diffraction point of view, the slits do not interact, cannot see each other and act almost undisturbed in the formation of the same transmission resonance provided by the single-slit configuration. In contrast, abrupt changes of the field localization are evident when vertical resonances interact with horizontal ones. These variations can be linked in straightforward fashion to the strong perturbations of the diffraction efficiencies expected around the Wood's anomalies. We observe, in particular, that an enhanced reflection state, the point M in Fig. 4(a), is very close to an enhanced transmission state, the point V on the same figure, and the light suppression effect has a non-negligible bandwidth of several tens of nanometers. In the figure the state V appears only 7 nm away from the M state, and its bandwidth is extremely small, of the order of one nanometer in this case. The Poynting vector vortexes in the V state (Fig. 5(d)) are the result of the interaction between the zero-order diffracted wave and the leaky surface wave excited by the grating at the output interface. It is clear from the magnetic field localization that a hybrid mode embracing the whole grating (the sub-wavelength slits and the metal-air interfaces) is resonating collectively, allowing the transfer of light to the other side of the grating. In fact, the reader should observe that the coupling between surface waves excited on each side of the grating can be achieved only through the slits, given that the silver layer is much too thick (300nm) to allow coupling through its bulk sections. A closer look at the electric fields clarifies this concept.

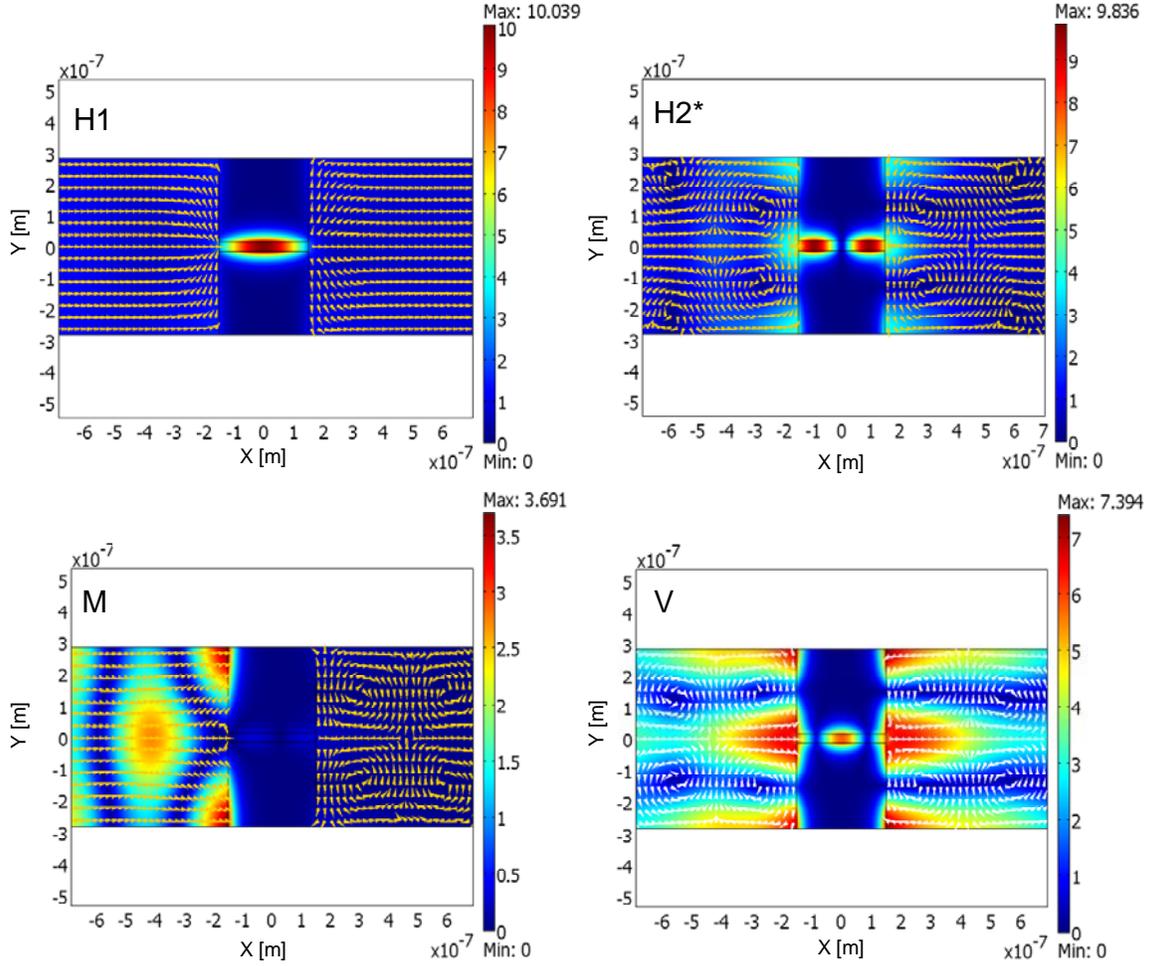

**Fig. 5:** (a) The color scale is the magnetic field amplitude at λ = 1.115 μm (point H1* in Fig. 3(c)) for an array with the same parameters described in Fig. 3(c) and Fig. 4(a), while the arrows sketch the normalized Poynting vector distribution; (b) Same as (a) at λ = 0.588 μm (point M in Fig. 4(a)); (c) Same as (a) at λ = 0.562 μm (point H2* in Fig. 4(a)); (d) Same as (a) at λ = 0.573 μm (point V in Fig. 4(a));

Figs. 6 depict the electric fields for the resonance V, demonstrating how both surfaces are involved in the resonant process, while the cavity mode couples the two surface waves.

The magnetic field localization of the H2* point (Fig. 5(b)), where both the input and output interfaces are significantly illuminated, reveals that the second-order horizontal resonance is strongly interacting with the vertical one: as a confirmation, vortexes similar to those excited in the V state are present almost at the same positions. Perhaps more interesting is the comparison between the V (Fig. 4(a)) and the H2* states (Fig. 5(b)). The magnetic field of the V state resembles an even, first order mode, and shows a sort of double localization, both inside the slit and along the metal walls, a display of the hybrid mode that includes cavity resonances and surface waves. On the other hand, the H2* state follows the localization of the second-order Fabry-Perot with two amplitude maxima cavity (odd mode). These field localization properties are very closely related to what occurs across the photonic band gap structure of a multilayer stack, where a π phase shift across the gap causes the electric and magnetic fields to trade place and roles.

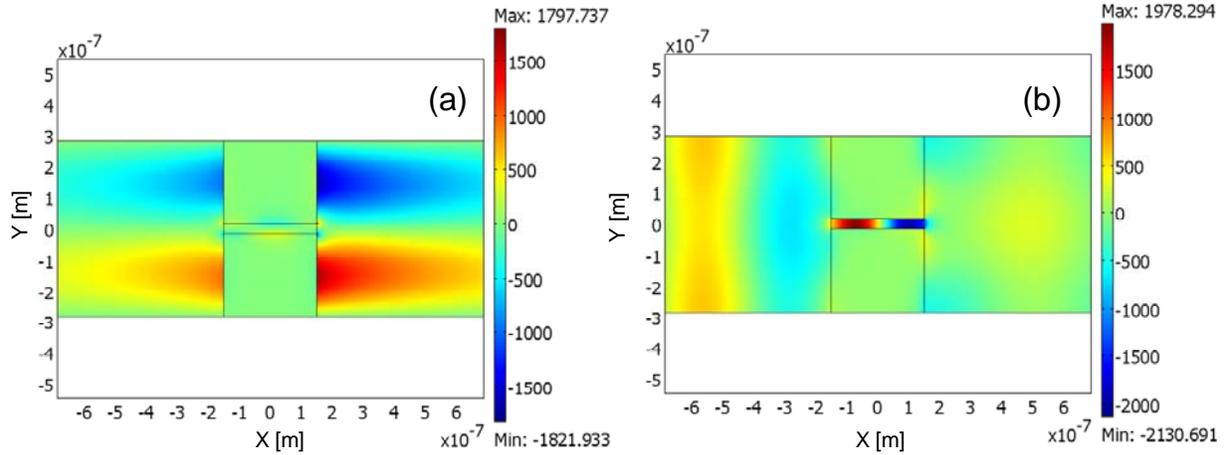

**Fig. 6:** (a) Longitudinal and (b) transverse electric field localization at λ = 0.573 μm (point V in Fig.5(d)); The electric field now interact with both the metal/air interfaces (a); It is evident from (b) how the coupling of the two surfaces is mainly due only to the resonance that takes place inside the sub-wavelength cavity. As a consequence, the V state is a result of a resonance involving the whole system and is not produced by surface waves alone.

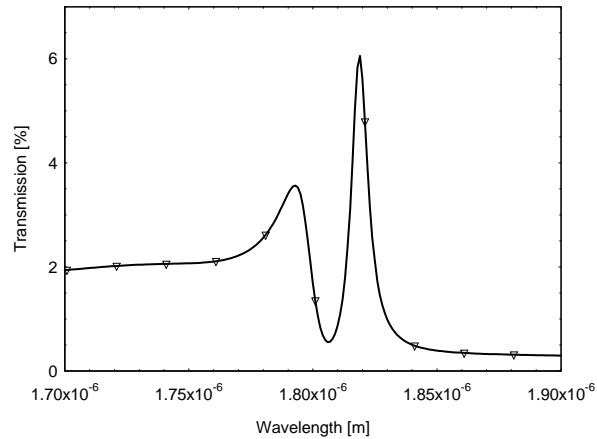

**Fig. 7**: Transmission of a grating made by the same slits of the gratings described in Fig.4 and pitch $p$=1800nm.

As already demonstrated for light transmission through single apertures surrounded by periodic grooves [34-36], leaky plasmons improve light coupling on the input interface and provide enhanced transmission and high broadside directivity at the exit interface of the grating. The same mechanism mediates the formation of the modes V and H2*. At the minimum point M, the incident light bends in front of the aperture and deflects back as if the grating were a smooth surface. The grating then acts as a perfect mirror, allowing the excitation of the unperturbed surface plasmon polariton with a minimum located in front of the slit. This causes a subsequent re-irradiation of light on the input side of the grating:

this virtually *chokes* the slits with a surface resonant state. Later we will show how this effect persists regardless of slit thickness and impinging wavelength, as long as the condition $\lambda_{sp} = p$ is satisfied. Absorption of ~2% and reflection of ~98% are consistent with values one expects from a smooth, free-standing silver interface. For completeness we also consider a case in which the first-order Wood's anomaly is isolated from the first-order horizontal resonance. The transmission (Fig. 7) shows a rudimentary gap located at the wavelength that excites the air/metal surface plasmon polariton.

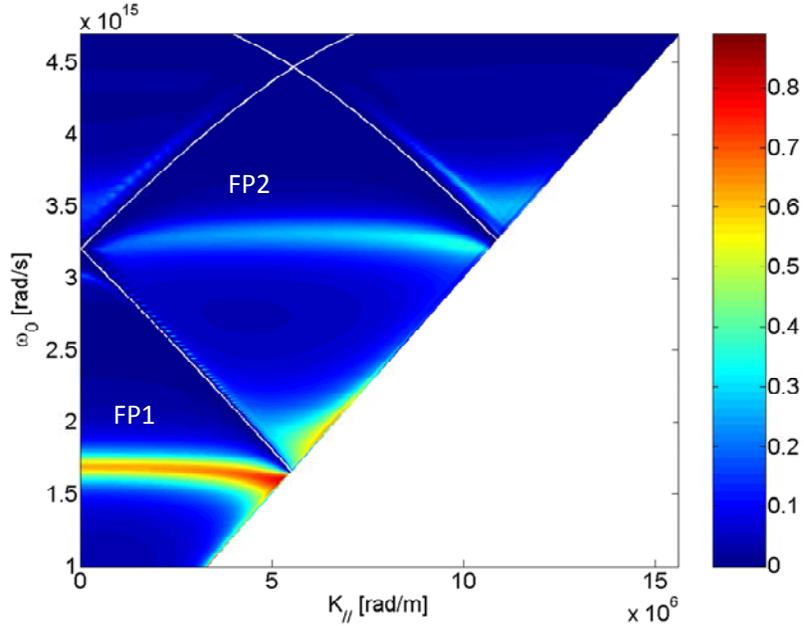

**Fig. 8**: Transmission map for an array of slits having the following geometrical parameters: $a$ = 32nm, $w$ = 300nm, $p$=566nm

**SURFACE PLASMONS AND WOOD'S ANOMALY: HOW ARE THEY LINKED?**

In the previous section we have shown how the formation of the band gap is directly linked to the periodicity of the structure and how the fields localize across the forbidden region. We have also demonstrated how to manage the interaction of surface waves alongside cavity resonances (Figs. 3), i.e. by appropriately tuning the array periodicity. However, an effective way to control such interaction is to tilt the structure and to scan the transmission at different incident angles. Unlike all the results reported so far, the transmission response in Fig. 8 is obtained this time by calculating the transmittance by varying the incident wavelength and the angle of the incident plane wave. The geometrical parameters of the array of Fig. 8 are the same as those of Figs. 4: $a$ = 32nm, $w$ = 300nm, $p$ = 566nm. We stress that the normal incidence transmission plotted in Fig. 4(a) is now sketched along the $k_{//}$=0 axis, and forms a band gap around the zero transmission state at $\lambda$ = 588nm ($\omega$=3.18×10$^{15}$ rad/s). At least two more observations may be inferred by looking at Fig. 8: i) by tilting the incident angle the band gap can cross the cavity mode orders of the system and locate the zero transmission state at different wavelengths within the Fabry-Perot cavity modes (FP1 or FP2 in Fig. 8), which are almost invariant with the incident angle; ii) the condition, for the excitation of the unperturbed (smooth, metal-air interface) surface plasmon, $k_{sp} = \left| k_0 \sin\theta + m\frac{2\pi}{p} \right|$, denoted by the white curves on the map, coincide systematically with the formation of a transmission minimum equal to the point M described in Fig. 4(a) and Fig. 5(b). However, the position of the band gap created in the spectrum is not invariant with aperture size, and helps to distinguish the role of the Wood's anomaly and SP.

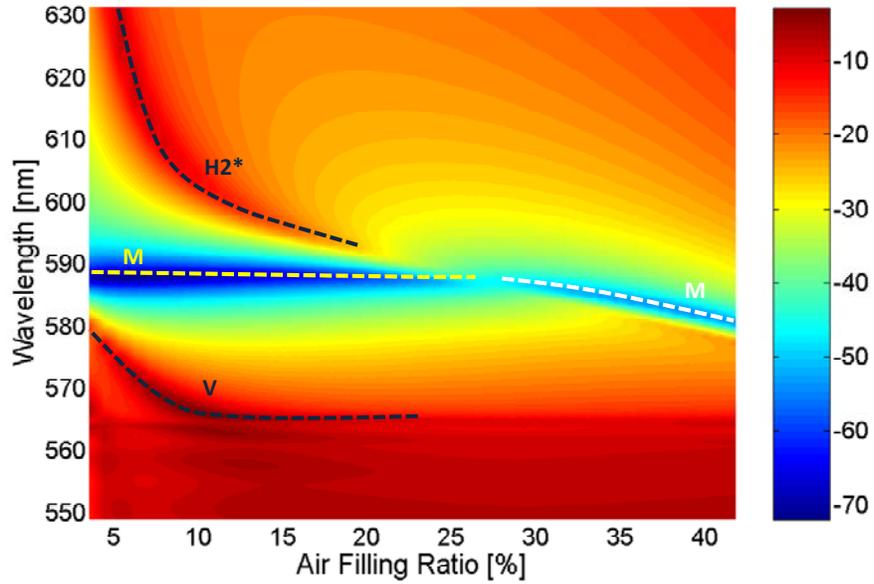

**Fig. 9**: Transmission map in logarithmic scale at normal incidence for a silver grating whose thickness is $w$ = 300 nm and $p$ = 566nm.

In Fig. 9 we report the transmission map on a logarithmic scale, obtained by varying incident wavelength and aperture size, for an array of slits carved on a 300nm-thick silver film. Once again we choose pitch size $p$ = 566nm. The incident wavelength is varied around the characteristic wavelength that excites a surface plasmon propagating at an effective wavelength equal to the pitch size, while the aperture size, or the air filling ratio, is varied from 20 nm (air filling ratio ~3%) to 560 nm (air filling ratio ~99%). Obviously total transmission approaches 100% by simply increasing the air filling ratio, since we are reducing the amount of metal in the structure. The most interesting feature of this map is the section highlighted by the yellow and white dashed lines, both marked with the label M. These lines correspond to the minimum transmission state described above, occurring at $\lambda_{sp}$ = $p$. According to the results presented in the previous section, the zero transmission state falls systematically at the wavelength where the grating wave vector $2\pi/p$ excites the unperturbed surface plasmon. As long as the air filling ratio increases and the structure becomes less metallic, three effects take place: (i) the effective refractive index of the fundamental TM mode in the slit waveguide decreases and tends to unity, so that all the Fabry-Perot resonances shift toward shorter wavelengths; (ii) the bandwidths of the Fabry-Perot resonances are broader since the grating tends to be transparent; *(iii) the condition for the minimum transmission state moves from the surface plasmon excitation (yellow line in Fig. 9) to the classic Rayleigh minimum condition (white line in Fig. 9), occurring at $\lambda = p$.* The transition from the plasmonic minimum (yellow line) to the purely photonic minimum (whit line) is evident by looking at the slope of the white line, starting from an air filling ratio of ~28%.

Let us now consider two different sections of Fig. 9: the first one corresponds to $\lambda_{sp} = p$ and the second one is $\lambda = p$, with $p$=566nm. The excitation of the bound surface wave having an effective wavelength equal to pitch size ($\lambda_{sp} = p$) causes a zero transmission point (black curve - circle markers in Fig. 10). By cutting the map at the Wood-Rayleigh anomaly condition $\lambda = p$, we obtain instead a modulated transmission efficiency (blue curve - square markers in Fig. 10). In this case the incident wavelength is closer to the narrow, vertical resonance condition, i.e. the V state described in Fig. 4(a) and 5(b), whose spectral location is slightly modified by the aperture size. Indeed, the V maximum state intercepts the Wood-Rayleigh anomaly condition for an air filling ratio of ~10%, i.e. an aperture size of ~57nm, as indicated by the transmission maximum in the blue curve of Fig.10.

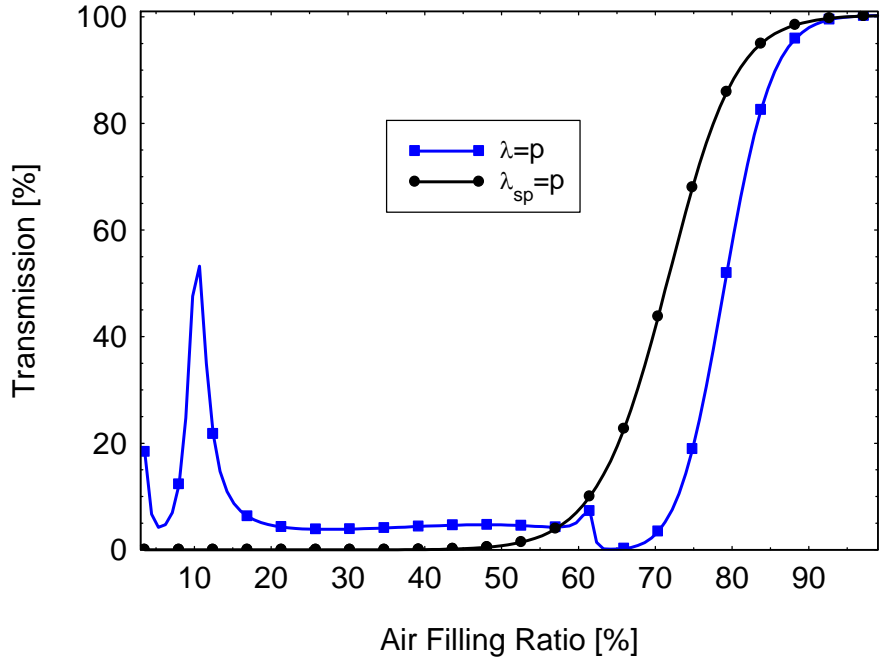

**Fig. 10**: Plot of two sections of Fig. 9 on a linear scale. The black line (circle markers) corresponds to an excitation of a surface wave matching the grating pitch ($\lambda_{sp} = p$). The blue line (square markers) is obtained cutting the map in Fig.9 at the points $\lambda = p$, that is the condition for Wood-Rayleigh anomalies.

The fact that the spectral position of the V mode is influenced by aperture size reveals the hybrid nature of this state. A re-examination of Fig. 5(d) confirms that even if the V mode is well localized at both air-metal surfaces, these two leaky surface waves are coupled via the fundamental propagating mode present inside the slit waveguide. The strength of this coupling is modulated by variations of the air filling factor, leading to a slight shift of the V mode resonance wavelength (see the black dashed line labeled V in Fig.9). A similar conclusion can be drawn on the blue-shift of the H2* mode: the transmission band corresponding to this hybrid mode shifts from 630nm at an air filling ratio of 5% to 600nm for an air filling ratio of 10%, and it disappears altogether for larger aperture sizes due to the crossing of the high reflection state occurring at $\lambda_{sp} = p$.

For the sake of completeness we also calculated the transmission response as a function of both impinging wavelength and metal thickness. Both aperture and pitch are constant and equal to $a = 32$nm and $p = 566$nm, respectively. In Fig.11 one may easily recognize some of the features of Fig.1 (b), which was obtained for a single aperture of the same size: the first and the second order Fabry-Perot modes yield the same spectral position and are altered only where they are crossed by the transverse resonating surface modes (horizontal section $\lambda_{sp} = p$ highlighted in white). Moreover, two regimes are distinguishable in this map, separated by the vertical dashed line. The region on the right refers to the thick film regime and is characterized by the interaction of longitudinal and transverse resonances that interact thanks to the guided modes of the slits. In contrast, in the left region, i.e. the thin film regime, the guided modes of the slits cannot resonate and the enhanced transmission is achieved only through evanescent wave coupling that takes place either inside the slits or the bulk metal. In particular in this regime the coupling of the input and output surface plasmons induces the formation of long and short range plasmons whose dispersion can be significantly distorted with respect to the relation $k(\omega) = \frac{\omega}{c}\sqrt{\frac{\varepsilon_1 \varepsilon_2}{\varepsilon_1 + \varepsilon_2}}$ [37].

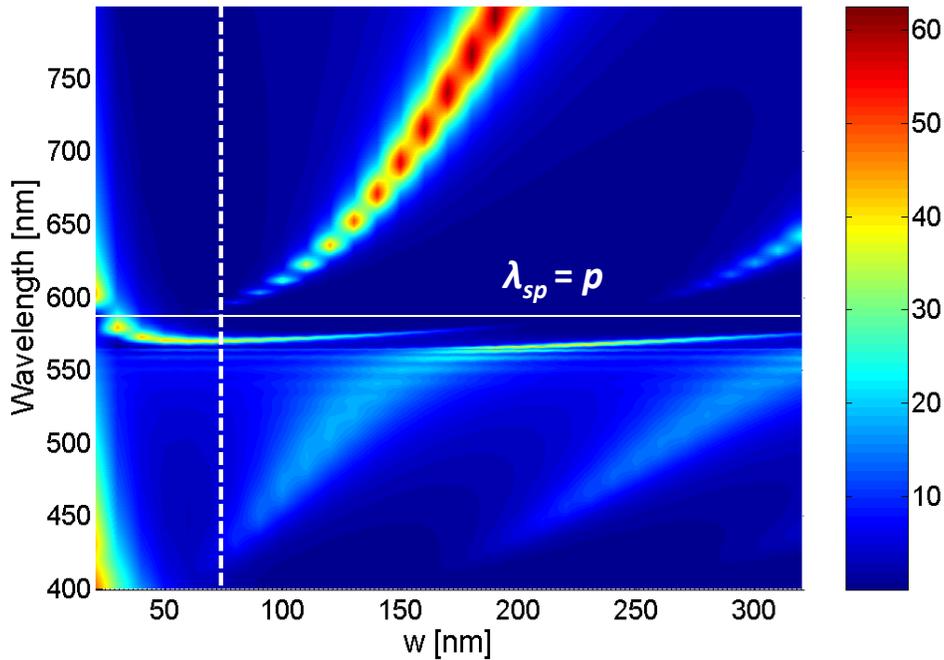

**Fig. 11**: Transmission map at normal incidence for a silver grating with variable thickness, pitch size $p = 566$nm and aperture size $a = 32$nm. The vertical dashed line traces an ideal boundary between the thick and the thin film regimes. The horizontal line corresponds to the condition $\lambda_{sp} = p$.

**FREE-STANDING AND SUPPORTED GRATINGS: THE ROLE OF SURFACE PLASMONS**

Let us now go back to the small aperture regime to explain further the actual role that surface plasmons have in these systems. As pointed out in both theoretical [18] and experimental results [20], the role of surface plasmons in the enhanced transmission regime is *negative*, in the sense that its excitation corresponds to a virtually zero transmission state, provided $\lambda_{sp} = p$. We have seen in the previous section how this condition persists for air filling ratios less than 50% (the metal dominates in the structure). Now we will show how this relationship is preserved regardless of the thickness of the metal film (Fig. 12) or the incident wavelength (Fig. 13). In Fig.12 we report the transmission coefficient for a plane wave tuned at λ = 584nm impinging at normal incidence on a grating made by a one-dimensional periodic arrangement of slits with aperture size *a* = 32nm. The map is obtained for variable film thicknesses and pitch sizes: the grating thickness *w* varies from 150nm to 1150nm (x-axis) and the pitch size *p* varies from 450nm to 2μm (y-axis). Pitch size has been normalized to the surface plasmon wavelength excited by the impinging wavelength, $\lambda_{sp}$ ~ 562 nm. Using this normalization, the transmission map unequivocally and clearly shows the recurrence of a persistent transmission minimum condition at multiple integers of the surface plasmon wavelength. In similar fashion, we may fix aperture size (*a* = 32nm) and the film thickness (*w* = 340nm), and vary incident wavelength and pitch size (Fig.13). This additional analysis confirms once again both the detrimental role of the unperturbed surface plasmons on transmission and the effects of the interaction of surface waves with Fabry-Perot resonances: three dark zones cross the resonances in Fig.13, creating a strong minimum in correspondence of pitch sizes multiples of $\lambda_{sp}$, where $\lambda_{sp} = \lambda / n_{sp}$, and $n_{sp} = \sqrt{\varepsilon_1 \varepsilon_2 / (\varepsilon_1 + \varepsilon_2)}$. Along with the zero transmission we note once again the opening of band gaps across the resonances.

Another important aspect of the transmission process in metal gratings is the influence of a supporting dielectric substrate. This, in fact, could be a major issue to be considered in experiments, since the realization of free-standing gratings could be very challenging. As our results above suggest, the interaction of surface waves with the grating is very important when their effective wavelengths approach multiples of the pitch size. This consideration remains valid also when a supporting dielectric material is in contact with the metal: two unmatched interfaces introduce not only an alteration of the in- and out-coupling of the nano-cavity, but will also support two different surface mode wavelengths according to their dielectric permittivity. Both these waves will play an identical role on the transmission process, regardless of the choice of the first interface encountered by the light. As an example, we report here the same map of Fig.12 for a metal grating with one surface exposed to air and the other one supported by a SiO$_2$ layer. For the sake of simplicity the SiO$_2$ layer is considered semi-infinite in the simulations, in order to avoid cavity effects and excitation of waveguide modes due to the finite thickness of the dielectric.

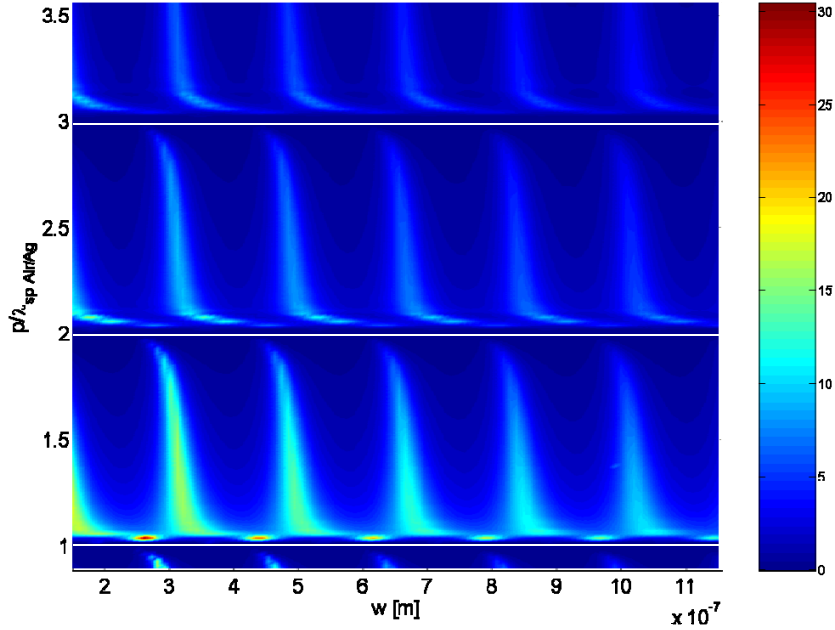

**Fig. 12**: Transmission map at normal incidence for a silver grating with apertures of 32 nm. The incident wavelength is $\lambda = 584$ nm and the excited surface plasmon is $\lambda_{sp} = 561$ nm. The map has three strong minima in correspondence of pitch sizes multiple of $\lambda_{sp}$, proving the negative role of surface plasmon in the transmission process.

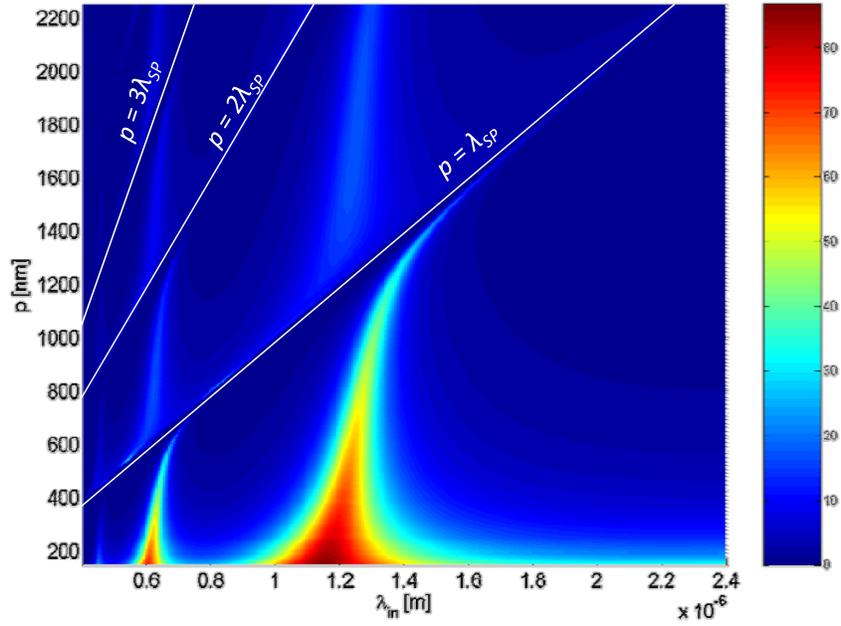

**Fig. 13**: Transmission map at normal incidence for a silver grating with apertures of 32 nm and thickness 340 nm. Three evident cuts in the Fabry-Perot resonances are reported when the pitch size matches $\lambda_{sp}$ or its multiples. The reshaping of the resonance and the formation of the band gap in the regions interested by that critical condition is also remarked in this map.

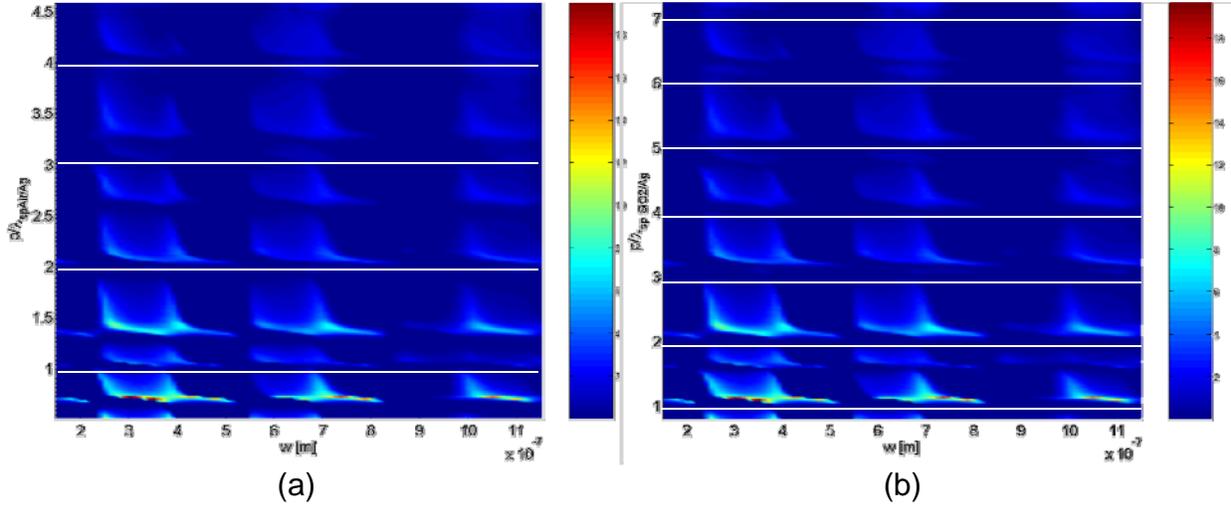

(a) (b)

**Fig. 14**: Transmission map at normal incidence for a silver grating with apertures of 32nm and with one surface exposed to air and one exposed to SiO$_2$ ($n_{SiO2}$ = 1.46). The incident wavelength is $\lambda$ = 584nm and the excited surface plasmons are respectively $\lambda_{sp\ Air/Ag}$ ~ 561 nm (for the Air/Ag interface) and $\lambda_{sp\ Ag/SiO2}$ ~ 310 nm (for the Ag/SiO$_2$ interface). The maps are identical and they have been normalized to (a) $\lambda_{sp\ Air/Ag}$ and (b) $\lambda_{sp\ Ag/SiO2}$, revealing how both these surface waves have the same detrimental role in the transmission process.

The light impinges at normal incidence on a grating of variable thickness and pitch size, having apertures $a$ = 32 nm wide. The plane wave used in the simulations is tuned at $\lambda$ = 584nm, meaning that the two surface waves generated at smooth Ag/Air and SiO$_2$/Air interfaces propagate respectively at $\lambda_{sp\ Air/Ag}$ ~ 562 nm (as in Fig. 12) and $\lambda_{sp\ SiO2/Ag}$ ~ 310 nm, assuming a refractive index $n_{SiO2}$ = 1.46 for SiO$_2$. Differently from the free-standing grating analysis (Fig. 12), in this case we have at our disposal two different normalizations for the pitch size $p$ (y-axis) to stress the role played by both surface waves: the same transmission map is plotted by normalizing the pitch size by $\lambda_{sp\ Air/Ag}$ in Fig.14 (a) and with respect to $\lambda_{sp\ Ag/SiO2}$ in Fig.14 (b). One may easily identify two sets of systematic zero transmission states for all the pitch sizes matching multiples of the unperturbed surface plasmon wavelengths on both sides of the film, giving us a precise idea of what to expect when we design or measure the transmission of a supported metal grating.

**ENHANCED TRANSMISSION FOR TE POLARIZATION**

Much effort has been devoted to understand and design enhanced transmission structures operating in different spectral ranges under TM polarized illumination, for the simple reason that surface plasmons have been considered an essential mechanism to enable extraordinary light transmission through metal layers. In one-dimensional metal gratings illuminated by TM-polarized light, as the arrays of slits analyzed in the present work, broad transmission peaks can be opened by the Fabry-Perot resonant effect acting on the fundamental TM-guided mode inside the slit. Nevertheless, if the aperture size is increased or the nano-slits are filled with a high dielectric permittivity material, so that TE-polarized light can couple to waveguide modes above their cut-off frequency, one can attain enhanced transmission for TE-polarized light in specific wavelength ranges by making these propagating modes resonant inside the slits, in the same way the principal TM mode of the slit waveguide is responsible for the horizontal cavity resonances reported in the previous sections. Using this concept un-polarized light can be channeled through a perforated metal layer provided that at least one mode for TE- polarization is above cut-off [38]. The other resonant mechanism that opens enhanced transmission regions is the excitation of vertical resonances, i.e. surface waves in which the entire grating resonates, so that part of the diffracted waves coupled to leaky modes supported by one side or the other tunnels through the grating. In particular, for TM polarization these leaky modes are the same leaky surface plasmons supported by the periodically perturbed metal-dielectric surface. Under TE illumination, leaky modes can be excited by adding a dielectric slab on one or both sides of the grating. The primitive reciprocal lattice vector can then match the wave vector of the modes guided by the metal-dielectric interface: this waveguide structure, formed by the dielectric slab surrounded by free-space and the periodically etched metal layer, supports both TE and TM surface modes.

In this kind of structures, where slits are carved on metal-dielectric stacks, the complexity is magnified by the simultaneous presence of pure grating effects (like the horizontal and vertical resonances reported above) and other phenomena linked to the response of flat metal-dielectric multilayers to propagating and evanescent modes, like photonic band gap effects [39] or super-focusing and super-resolution effects [40,41]. In this context, it is may be useful to classify several resonant mechanisms, which can be grouped in the following five categories: 1) horizontal resonances provided by the guided modes in the slits, always present for TM light and conditionally present for TE light above the cut-off frequency of the fundamental TE mode; this mechanism is present for both free-standing gratings and gratings loaded with dielectric slabs, and they are present also in a single-slit structure; 2) horizontal resonances involving the bulk metal layer, and enabled by a proper choice of the optical length of the surrounding dielectric layers; these structures are known as induced-transparency filters [42,43], do not require the grating to provide enhanced transmission, and work for both TE and TM polarizations; 3) Fabry-Perot, horizontal resonances when two or more metal layers are present; in this case the structure, which resembles a photonic band gap (PBG) stack, acts like a transparent metal, since it can open wideband transparency regions independent of incident polarization; 4) vertical resonances excited by leaky surface waves, TM polarized, supported also by a simple free-standing grating; 5) vertical resonances linked to leaky modes supported by dielectric slabs loading the metal grating, which are present for TE and TM polarization.

To highlight all these complex, varied phenomena we have considered the grating sketched in Fig. 15(a). This structure consists of a silver grating 75 nm-thick pierced with rectangular slits 50 nm-wide and a pitch size of 566 nm, while two dielectric slabs 80 nm-thick with a refractive index of 3.47 (the material could be a high-index semiconductor as Si or GaAs) are arranged symmetrically on both sides of the grating. We stress that the optical behavior of the flat, three-layer structure, without considering the grating effect of the slits, is not trivial as might seem at a first superficial analysis. This three-layer device, indeed, embodies a typical induced transmission filter [42], having a low-reflection region around 450 nm, as depicted in Fig. 15(b). In this particular case, the portion of energy not reflected back by the multilayer is completely absorbed by the evanescent waves inside the metal. A reduction of the metal thickness will further frustrate reflection in the same bandwidth, while the transmission/absorption ratio will increase. The introduction of the slits in this structure, for TE polarized light, induces extremely narrow transmission and absorption maxima directly related to vertical, guided (but leaky) modes propagating along the input and output slabs. In this system the nano-slit does not support any horizontal, propagating mode, since the main TE mode is below cut-off for the wavelength range we are considering (400÷800 nm), so that the coupling between the input and output vertical modes of the surrounding slabs is purely evanescent. Transmission will be further enhanced, both in the induced transmission region and in the guided-resonance peaks, by thinning the silver layer, or by opening the slit width. We emphasize that this structure, under TE polarization, has an optical behavior similar to a metal grating made by circular holes illuminated with TM polarized light; in the latter case, the vertical modes are leaky surface plasmons. The spectra for TM polarization are illustrated in Fig. 15(d); in this case, the waveguide modes of the input and output slabs are coupled by the fundamental TM mode available in the central slit. For both polarizations, the position of the absorption and transmission peaks is tuned by the refractive index of the dielectric (or semiconductor) slabs, as well as by their thickness. It is clear that a proper design of the geometrical parameters of the grating, can lead to enhanced transmission for both TE and TM polarizations, at least in narrow bandwidths. This study requires a careful analysis of the dispersion properties of the structure, and it will be an interesting subject for future work.

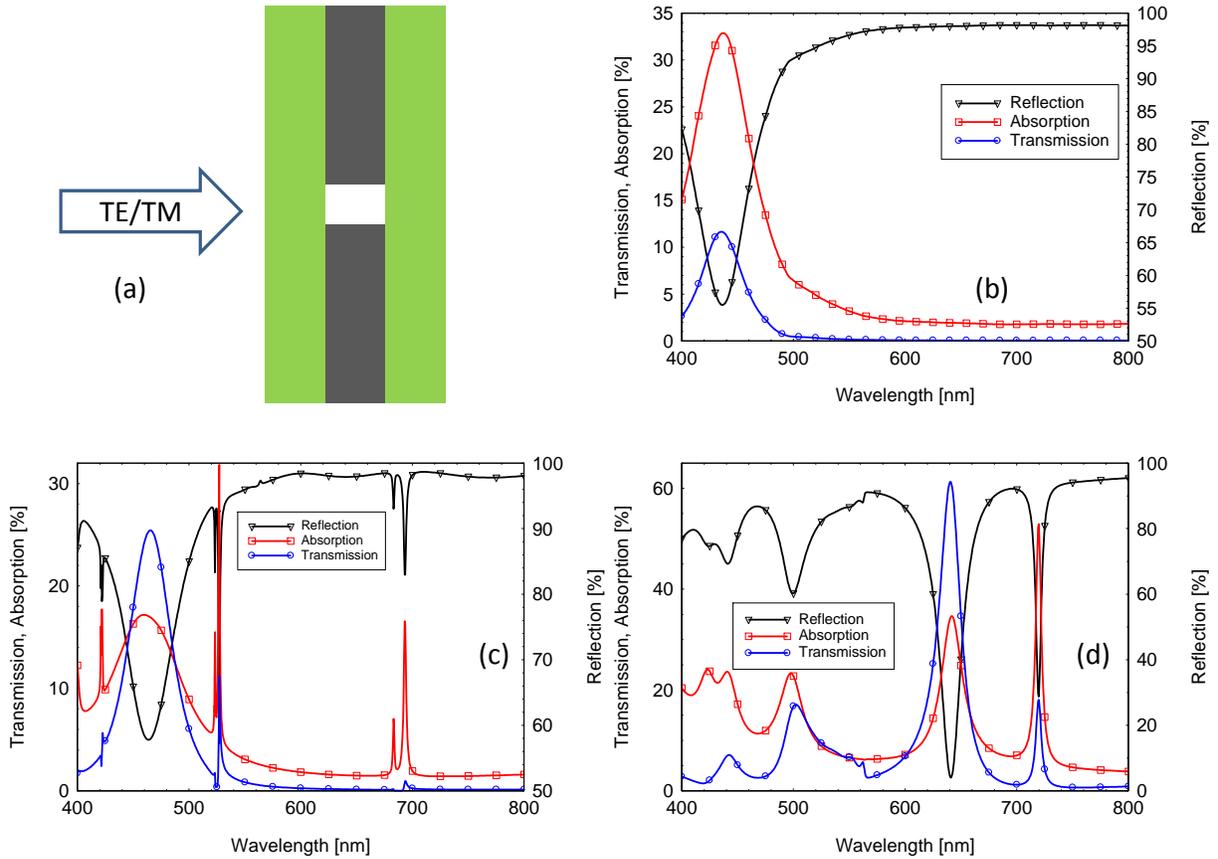

**Fig. 15**: (a) Scheme of the one period of the grating illuminated with TE and TM polarized plane waves at normal incidence. The green regions are two slabs 80nm-thick of a loss-less material with refractive index n=3.47. The silver grating (grey, central layer) has slits with a=50nm, while its thickness is w=75nm. The pitch size is p=566nm; (b) Transmission, absorption and reflection of the three-layer structure in (a) without apertures in the metal layer. (c) Spectra of the grating (a) under TE polarization. (d) Spectra of the grating (a) under TM polarization.

**Conclusions**

In summary we have analyzed several fundamental properties of enhanced transmission in metal gratings, showing the possibility to achieve full control of transmission efficiency. The combined study of a single slit system and its periodic counterpart reveals virtually identical spectral features, except at wavelengths that nearly match the grating pitch. In particular, we have identified the excitation of the unperturbed surface plasmon as the mechanism that inhibits light flow throughout the slits. This result has been confirmed on both free-standing gratings and structures supported by a dielectric substrate, providing an excellent guide tool for experiments. On the other hand, leaky surface waves excited by the grating act like vertical resonating modes, that are able to collect the impinging light and interact with horizontal, Fabry-Perot like modes, to form hybrid modes and trigger high transmission states. The hybrid nature of these modes has been confirmed by observing the field localization properties, and by mapping the behavior of the structure as a function of the air filling ratio, revealing their strong dependence on the aperture size. We also demonstrated that the excitation of surface plasmons is not a stringent requirement to achieve enhanced transmission regimes: embedding the grating within dielectric or semiconductor layers enables access to narrow bandwidth, high-transmission states even for TE polarization.